\documentclass[%
 reprint,
 amsmath,amssymb,
 aps,
]{revtex4-2}

\usepackage{graphicx}
\usepackage{amsmath}
\usepackage{dcolumn}
\usepackage{bm}
\usepackage[caption=false,font=normalsize,labelfont=sf,textfont=sf]{subfig}

\begin{document}

\preprint{APS/123-QED}

\title{Effect of antisite disorder on the magnetic and transport properties of a quaternary Heusler alloy}

\author{Srishti Dixit$^{1}$}
\author{Swayangsiddha Ghosh$^{1}$}
\author{Sanskar Mishra$^{2}$}
\author{Nisha Shahi$^{3}$}
\author{Prashant Shahi$^{2}$}
\author{Sanjay Singh$^{3}$}
\author{A. K. Bera$^{4}$}
\author{S. M. Yusuf$^{4}$}
\author{Yoshiya Uwatoko$^{5}$}
\author{C.-F. Chang$^{6}$}
\author{Sandip Chatterjee$^{1}$}
\email{schatterji.app@itbhu.ac.in}

\affiliation{$^{1}$Department of Physics, Indian Institute of Technology (Banaras Hindu University), Varanasi 221005, India}
\affiliation{$^{2}$Department of Physics, Deen Dayal Upadhyaya Gorakhpur University, Gorakhpur, India}
\affiliation{$^{3}$School of Material Science and Technology, Indian Institute of Technology (Banaras Hindu University), Varanasi 221005, India}
\affiliation{$^{4}$Solid State Physics Division, Bhabha Atomic Research Centre, Mumbai 400085, India and Homi Bhabha National Institute, Anushaktinagar, Mumbai 400094, India}
\affiliation{$^{5}$Institute for solid State Physics, University of Tokyo, Kashiwa, 2778581, Chiba, Japan}
\affiliation{$^{6}$Max Planck Institute for Chemical Physics of Solids, Nöthnitzer Str. 40, 01187 Dresden, Germany}






\begin{abstract}
Spin gapless semiconductors based Heusler alloys are the special class of materials due to their unique band structure, high spin polarization and high Curie temperature. These materials exhibit a distinct electronic structure: a nonzero band gap in one spin channel while the other spin channel remains gapless, making them highly suitable for tunable spintronics. In this study, a comprehensive analysis of structural, magnetic, thermoelectric, and transport properties of the quaternary Heusler alloy CoFeMnSn is conducted. X-ray diffraction and Neutron diffraction analyses confirm a well-ordered structure with partial antisite disorder between Co $\leftrightarrow$ Fe and Mn $\leftrightarrow$ Sn atoms. Magnetic studies show that the material exhibits room-temperature ferromagnetism, with a Curie temperature of around 660 K. Notably, we observe an anomalous Hall effect (AHE) linked to intrinsic mechanisms driven by Berry curvature, underscoring the intricate relationship between structural disorder and electronic behavior. Transport measurements also highlight the impact of antisite disorder on the systems, with resistivity decreasing as temperature increases. These insights position CoFeMnSn as a promising material for future spintronic devices and advanced technological applications.
\end{abstract}

\maketitle 


\section{Introduction}
 The Heusler family comprises a wide range of captivating magnetic materials, including half-metallic ferromagnets \cite{bainsla2015high}, spin-gapless semiconductors \cite{bainsla2015origin}, bipolar magnetic semiconductors \cite{nag2021bipolar}, and spin semimetals \cite{venkateswara2019coexistence}. These materials possess a stable structure, high spin polarization, and high ordering temperatures, which sets them apart from other spintronic materials and makes them well-suited for many applications. According to their atomic arrangements, different types of systems are possible: Half Heusler (XYZ), full Heusler (X$_2$YZ), and quaternary Heusler alloys (QHAs) (XX’YZ).  The X, X’, and Y belong to the transition metal elements and Z belongs to the p-block element. Among all the possible structures of Heusler alloys QHAs are composed of four different elements (XX’YZ) \cite{graf2011simple}, offering a rich playground for tailoring their physical and chemical characteristics. The presence of two distinct sublattices in QHAs imparts numerous advantages, making these materials highly intriguing for scientific exploration and technological applications. One of the key advantages is that the electronic band structure, spin polarization, and magnetic moments can be selectively adjusted by manipulating the composition and arrangement of elements on the two sublattices. Among all the QHAs, the Co-based Heusler alloys have become a pivotal focus in research, driven by their notable attributes, including high Curie temperatures $(T_C)$, remarkable spin polarization, and the ability to fine-tune electronic structures \cite{nag2022cofevsb,mouatassime2021magnetic,dixit2023existence}. Furthermore, these alloys typically exhibit a substantial anomalous Hall effect (AHE) \cite{xia2022magnetic}.

The normal Hall effect is defined as when an external magnetic field is applied perpendicular to the current flow in a conductor, resulting in the formation of a transverse voltage called Hall voltage \cite{vsmejkal2022anomalous,nagaosa2010anomalous}. The anomalous Hall effect (AHE) is an observable phenomenon in materials that have broken time-reversal symmetry where an extra contribution to the transverse voltage is seen. The emergence of the AHE may be attributed to the interaction between magnetization and spin-orbit coupling. The basic understanding of the observation of AHE suggests that it arises from the extrinsic and intrinsic mechanisms \cite{tian2009proper,karplus1954hall,onoda2006intrinsic}. The extrinsic mechanism involves skew scattering and side-jump processes, where skew scattering results in asymmetric deflection, and side-jump leads to a lateral shift in the path of spin-polarized charge carriers.  \cite{berger1970side}. In contrast, the intrinsic mechanism is linked to the Berry curvature in momentum space, which is a property of the electronic band structure. Notably, certain Co-based compounds within this category, such as CrFeVGa \cite{nag2023nontrivial} and CoFeVSb \cite{nag2022cofevsb}, have been reported to display AHE associated with Berry curvature \cite{roy2020anomalous}. Interestingly, the skew scattering mechanism tends to be less prominent in these alloys, as it often makes a negligible contribution to the AHE. This emphasizes the need for a comprehensive understanding of the AHE in cobalt-based Heusler alloys, as it holds significant implications for their potential applications in spintronics and related technologies \cite{alijani2011electronic}. 

However, Heusler compounds, particularly QHAs, are susceptible to antisite disorder within the d-block elements, a characteristic accentuated by the presence of three different d-block elements in the formula. Antisite disorder, characterized by the substitution of atoms from one type for another within the crystal lattice, is a prevalent defect in Heusler alloys and can significantly influence their electronic, magnetic, and thermal properties \cite{shen2020local,vilanova2011influence,hazra2018effect}. Notably, in QHAs, antisite disorder is strategically employed to tailor the electronic, magnetic, and transport properties of the system \cite{malik2022antisite}. Previous studies on Co-based magnetic Heusler compounds initially suggested that antisite disorder might be unfavorable for increasing the AHE, such as in thin films of Co$_2$MnGa and Co$_2$MnAl exhibited smaller $\sigma_{xy}^A$ values compared to well-ordered single crystals \cite{sakuraba2020giant,markou2019thickness}. However, recent investigations have unveiled an increased intrinsic AHC in Fe-based \cite{mende2021large}, which is attributed to antisite disorder in the system. Furthermore, in the Co$_2$FeAl Heusler compound, antisite disorder between transition metals elements (Fe and Al)  led to enhancement in intrinsic AHC. These findings underscore the nuanced role of antisite disorder in shaping the properties of Heusler alloys and offer avenues for tuning in advanced applications.

In the existing literature, the CoFeMnSn (CFMS) system has been identified as a half-metallic ferromagnet (HMF) through a combination of experimental and theoretical data \cite{xia2023structural}. Another independent investigation by Gupta \textit{et. al.}\cite{gupta2023spin} supports the spin-gapless semiconducting (SGS) nature of the system, both theoretically, experimentally and through transport properties analysis. Notably, while the former study verifies the system's structural purity, the later reveals a slight deviation from stoichiometry. Additionally, the later study explored the impact of pressure on CFMS, observing a transformation from SGS to HMF under pressure. In this present study, we delved into the comprehensive characterization of the quaternary Heusler alloy system CFMS, employing structural, magnetic, and transport measurements. Our structural analysis confirms the presence of antisite disorder in the system. Further exploration focuses on understanding the influence of antisite disorder on the magnetic and transport properties of the system. Magnetization studies affirm the ferromagnetic nature of the CFMS system, while transport data reveals the presence of AHE attributed to an intrinsic mechanism associated with Berry curvature. This multifaceted investigation contributes valuable insights into the intricate interplay between structural disorder and the magnetic and transport behaviors of quaternary Heusler alloys.

\section{Experimental Details}
The Arc melting method is used to make the polycrystalline CFMS sample. The constituent elements with a purity of 99.95\% were taken in a stoichiometric ratio, i.e., 1:1:1:1. The sample was melted several times to ensure homogeneity, and then the melted ingot was sealed in a quartz tube under vacuum. The sealed tube is placed into the furnace at 1073 K for five days for the annealing process and then followed by the quenching process in the ice water. The powder X-ray diffraction (XRD) of the system is performed after crushing the sample into the fine powder. The powder XRD was done using the Rigaku MiniFlex II DESKTOP powder diffractometer (Cu  $K_{\alpha}$ radiation, $\lambda = 1.54184 \, \text{Å})$. The Rietveld refinement of the X-ray spectrum is done using the  Fullprof software to determine the lattice parameter and atomic position of the system. For the transport study, a 1.28 mm $\times$ 0.62 mm $\times$ 0.23 mm rectangular piece is cut with a diamond cutter. The physical properties of the system are measured by the Physical Properties Measurement System (Quantum Design). For longitudinal resistivity ($\rho_{xx}$) four probe method is used for which the probe was made using the silver paste and gold wires. The current of magnitude 3 mA is applied in the X- X-direction and voltage is also measured in the X-direction. For magneto-transport the current is applied along X- direction and magnetic field (B) is applied perpendicular to the sample let say Z- direction and voltage is measured along the Y- direction. The Magnetic measurements of the system were performed using a superconducting quantum interference device-based magnetometer (MPMS, Quantum Design). Room temperature neutron diffraction pattern was recorded using the five linear position sensitive He3 detector-based PD-2 diffractometer ($\lambda$ = 1.2443 Å) at Dhruva reactor, Trombay, INDIA, over a wide angular range over 5-140 deg. The temperature-dependent neutron diffraction patterns were recorded at 3, 25, 50, 100, 150, 200, and 300 K by using the three linear position-sensitive He3 detector-based powder diffractometer PD-1 ($\lambda$ = 1.094 Å) at Dhruva reactor, Trombay, INDIA \cite{paranjpe1989neutron,paranjpe2002neutron}. For the neutron diffraction measurements, samples were filled in vanadium can of diameter 6 mm and attached to the cold head of a closed cycle helium refrigerator. Each of the patterns is measured for ~ 8.5 hours. The neutron diffraction patterns were analysed by the Rietveld method using the FULLPROF suite computer program \cite{rodriguez1993recent}. 

\section{Results and Discussions}
\subsection{XRD study}
\begin{figure}[ht]
\centering
\includegraphics[width=0.99\linewidth]{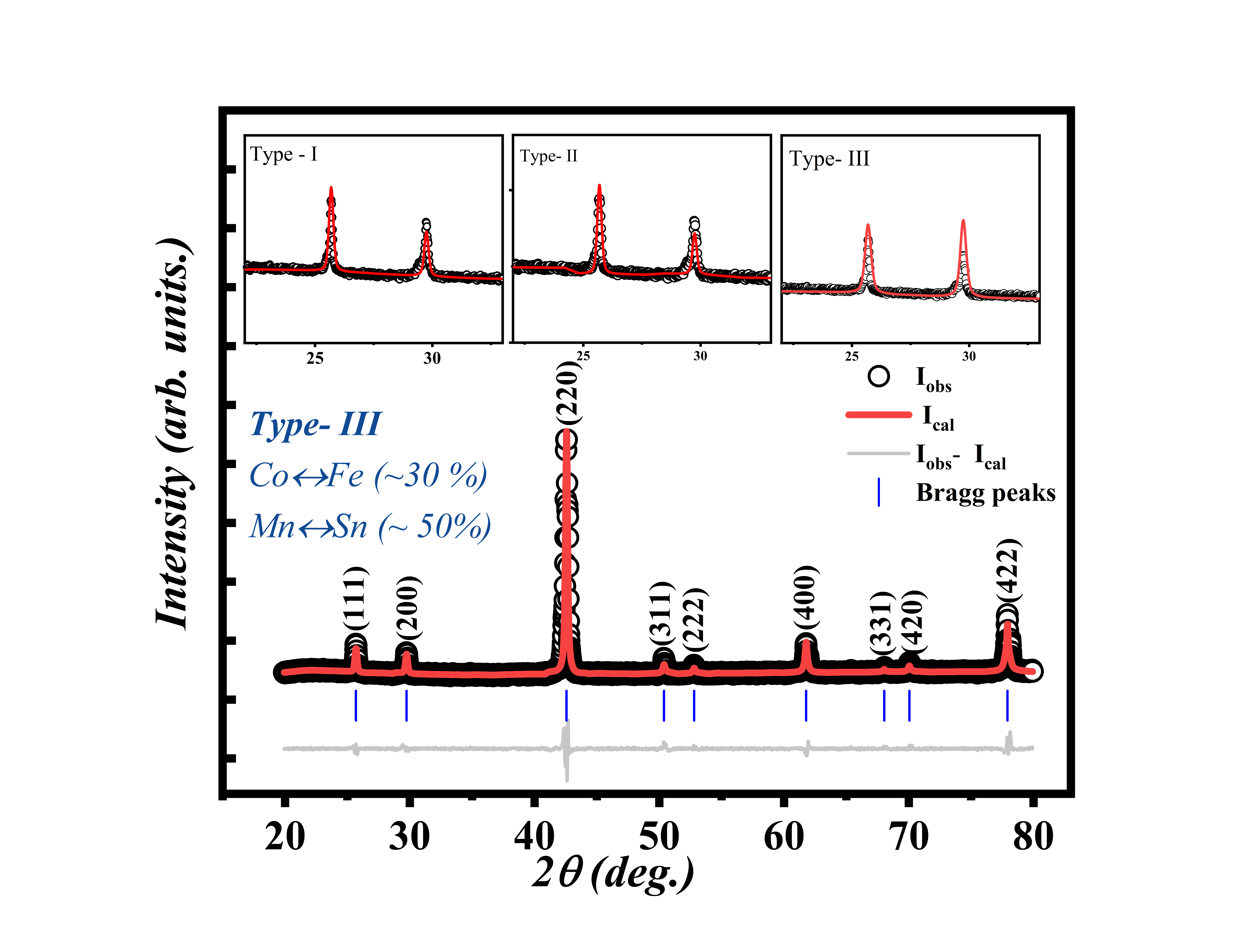}
\caption{Room temperature XRD pattern of CoFeMnSn along with the refinement for the Type III configuration with antisite disorder between Co and Fe, and Mn and Sn. (in the main graph). Inset is showing the Rietveld refinement of the XRD patterns for different configurations without the disorder.}
\label{fig:cd101}
\end{figure}

The crystal structure of bulk samples was determined through X-ray diffraction (XRD) measurements. Figure \ref{fig:cd101} displays the powder XRD of the CFMS system at room temperature (300 K).  The QHA CFMS crystallizes in the LiMgPdSn prototype structure with space group F$\overline{4}$3m. The observed XRD revealed clear characteristic peaks (220), (400), and (422) indicative of the cubic Heusler phase. Importantly, no diffraction peaks from the impure phase are detected, ensuring the purity of the samples. Additionally, distinct superlattice diffraction peaks (111) and (200) are prominently featured in the XRD pattern, suggesting a notable degree of atomic ordering within these alloys. For refinement, three non-degenerate structures were considered in the case of quaternary ($XX'YZ$) Heusler alloys, as given in Table I. The structure factor formula for the quarternary Heusler alloy is formulated as follows:
\begin{multline}
F_{hkl} = 4 \left( f_z + f_y e^{i\pi (h+k+l)} + f_x e^{\frac{i\pi}{2}(h+k+l)} \right. \\
\left. + f_x e^{-\frac{i\pi}{2}(h+k+l)} \right)
\end{multline}

Therefore,
\begin{equation}
    F_{111} = 4[(f_z-f_y ) -i(f_x - f_x)]
    \label{602}
\end{equation}
\begin{equation}
     F_{200} = 4[(f_z + f_y ) - (f_x + f_x)]
     \label{603}
\end{equation}
\begin{equation}
    F_{220} = 4[(f_z + f_y ) + (f_x + f_x)]
    \label{604}
\end{equation}

\begin{table}[ht]
\caption{Details of the Wyckoff positions for the atomic arrangement in the LiMgPdSn-type structure of the quaternary Heusler alloy CoFeMnSn.}
\begin{ruledtabular}
\begin{tabular}{ccccc}
CoFeMnSn & 4a & 4b   & 4c  & 4d  \\ 
 & (0,0,0) &  (1/2,1/2,1/2)  &  (1/4,1/4,1/4) &  (3/4,3/4,3/4) \\ \hline
Type I& Co & Fe & Mn & Sn \\
Type II & Co & Mn & Fe & Sn\\
Type III & Fe & Mn & Co & Sn \\
Type III  & Co/Fe & Co/Fe & Mn/Sn & Mn/Sn \\
 (disordered)  \\
\end{tabular}
\end{ruledtabular}
\end{table}

From the atomic structure formulae mentioned above in equations ($\ref{602}$), ($\ref{603}$), and ($\ref{604}$), it was clear that the prominent presence of peaks (111) and (200) arises due to the antisite disorder between the $Co\leftrightarrow Fe$ and $Mn\leftrightarrow Sn$ respectively. The refinement of the system is done with type III after including the approximate 30 \%  and 50 \% antisite disorder between $Co\leftrightarrow Fe$ and $Mn\leftrightarrow Sn$ atoms, respectively. The lattice constant for CFMS bulk was determined to be 6.0016 Å. 

\subsection{SEM Analysis}
\begin{figure}[ht]
\centering
\includegraphics[width=1.0\linewidth]{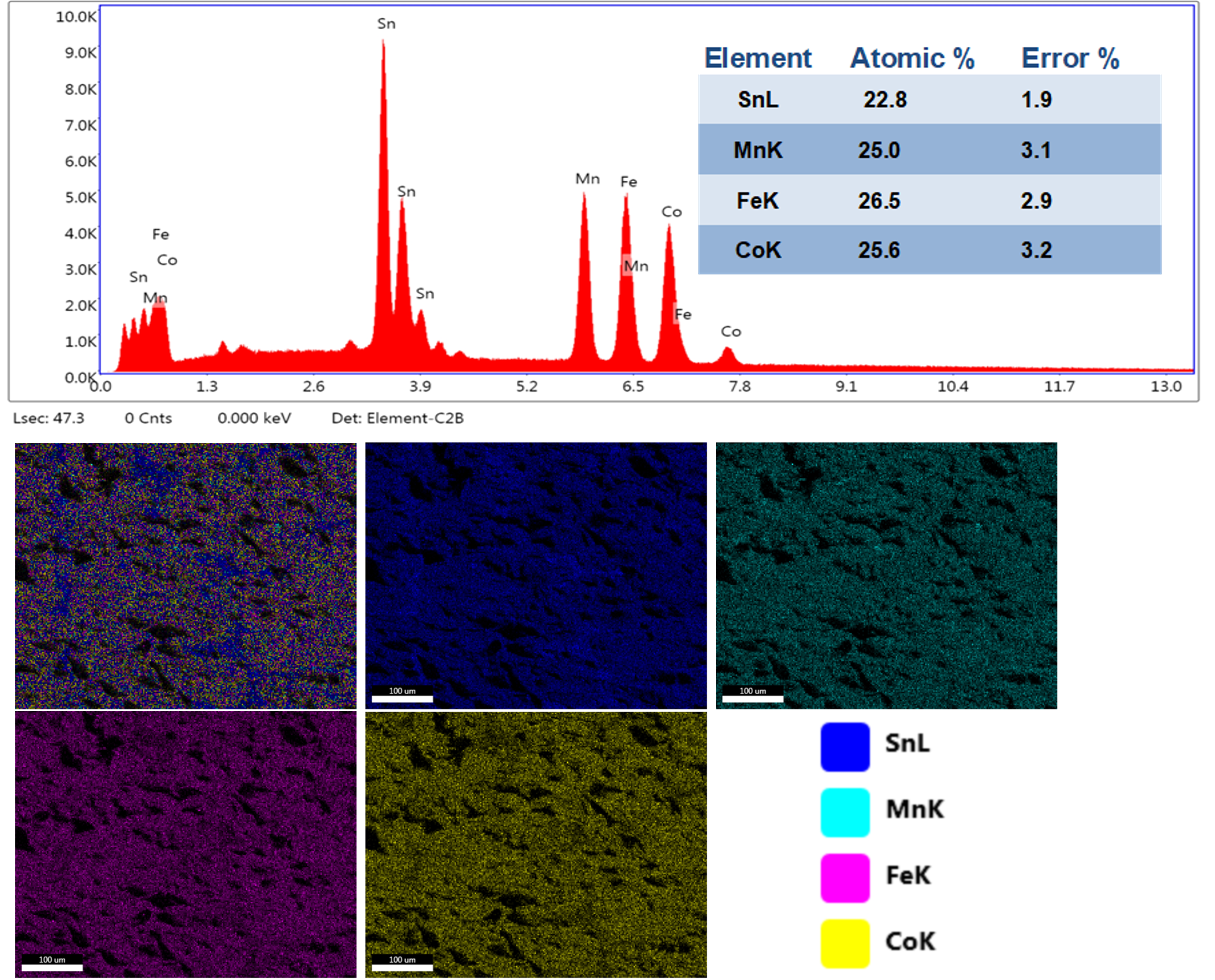}
\caption{Upper panel shows the EDS spectrum of the constituent elements of CoFeMnSn ( Inset shows the atomic percentage of respective elements) lower panel shows the composition mapping of constituent elements of the compound.}
\label{fig:cd602}
\end{figure}

The energy dispersive spectra (EDS) for the CFMS sample, along with the composition maps that were obtained from the EDS images, are shown in figure \ref{fig:cd602}. The atomic percentages of the constituent elements are closely matched with the stoichiometric of the compound, viz., 1:1:1:1, within the experimental error of not more than  3 $\%$. 

\subsection{X-ray absorption spectroscopy (XAS)} 
\begin{figure}[ht]
\centering
\includegraphics[width=1.0\linewidth]{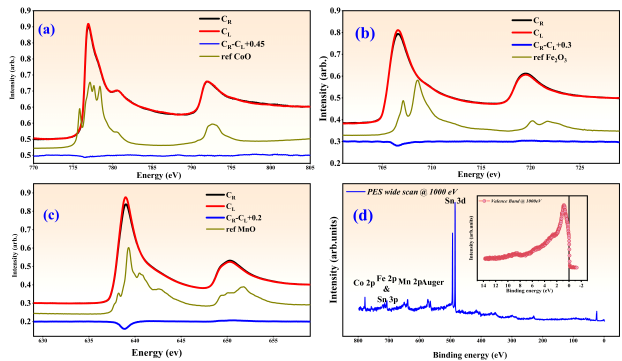}
\caption{The x-ray absorption spectra of Mn (a), Co (b), and Fe (c) along with the XMCD spectra of respective elements at 40 K (d) Photoelectron spectroscopy (PES) of the compound at 1000 eV, inset is showing the valance band scan of the system.}
\label{fig:cd603}
\end{figure}

The electronic characteristics of the CFMS sample were investigated by conducting X-ray Absorption Spectroscopy (XAS) and X-ray Magnetic Circular Dichroism (XMCD) measurements on the Co, Fe, and Mn elements at 40 K. The XAS/XMCD spectra of Co,Fe, and Mn-$L_{2,3}$, the edge is shown in figure \ref{fig:cd603}. The XAS spectrum of Mn shows the presence of two main peaks located at binding energies 638 eV and 651 eV. The absence of any multiplets in the $L_3$ and $L_2$ edge of the XAS spectra suggests the presence of the metallic nature of the elements in the system. In order to regain the information regarding magnetic ordering in the system, the XMCD signal of the elements is also shown in figure \ref{fig:cd603}.  
A negative dichroism signal was seen in the $L_3$ area, whereas a positive signal was observed in the $L_2$ region for Mn, Co, and Fe (see figure \ref{fig:cd603}. The presence of a non-zero signal confirms the magnetic ordering of the elements. The observed disparity between the spectra suggests that the system did not undergo an oxidation state, indicating its pure nature. Additionally, in the XAS/XMCD spectra of Co, which is recorded at  40 K.  The Co $L_{2,3}$ edge spectrum shows the presence of two main peaks at 777 eV and 793.5 eV. A shoulder peak is also observed at 3 eV higher binding energy. These structures correspond to the Heusler alloys due to the Co-Co bonding states within the molecular orbital calculations\cite{telling2006interfacial,elmers2003element}. The XAS/XMCD spectra have revealed that the magnetic moments of Mn, Co, and Fe are aligned in the same direction, hence confirming the ferromagnetic character of the system. 
Furthermore, The PES wide scan at 1000 eV was also performed for the system and it suggests the presence of constituent elements only. The inset of Figure \ref{fig:cd603} (d) is showing the valance band spectra of the system and it showed that finite states are present at the Fermi level.

\subsection{Neutron Diffraction Study}
\begin{figure}[ht]
\centering
\includegraphics[width=1.0\linewidth]{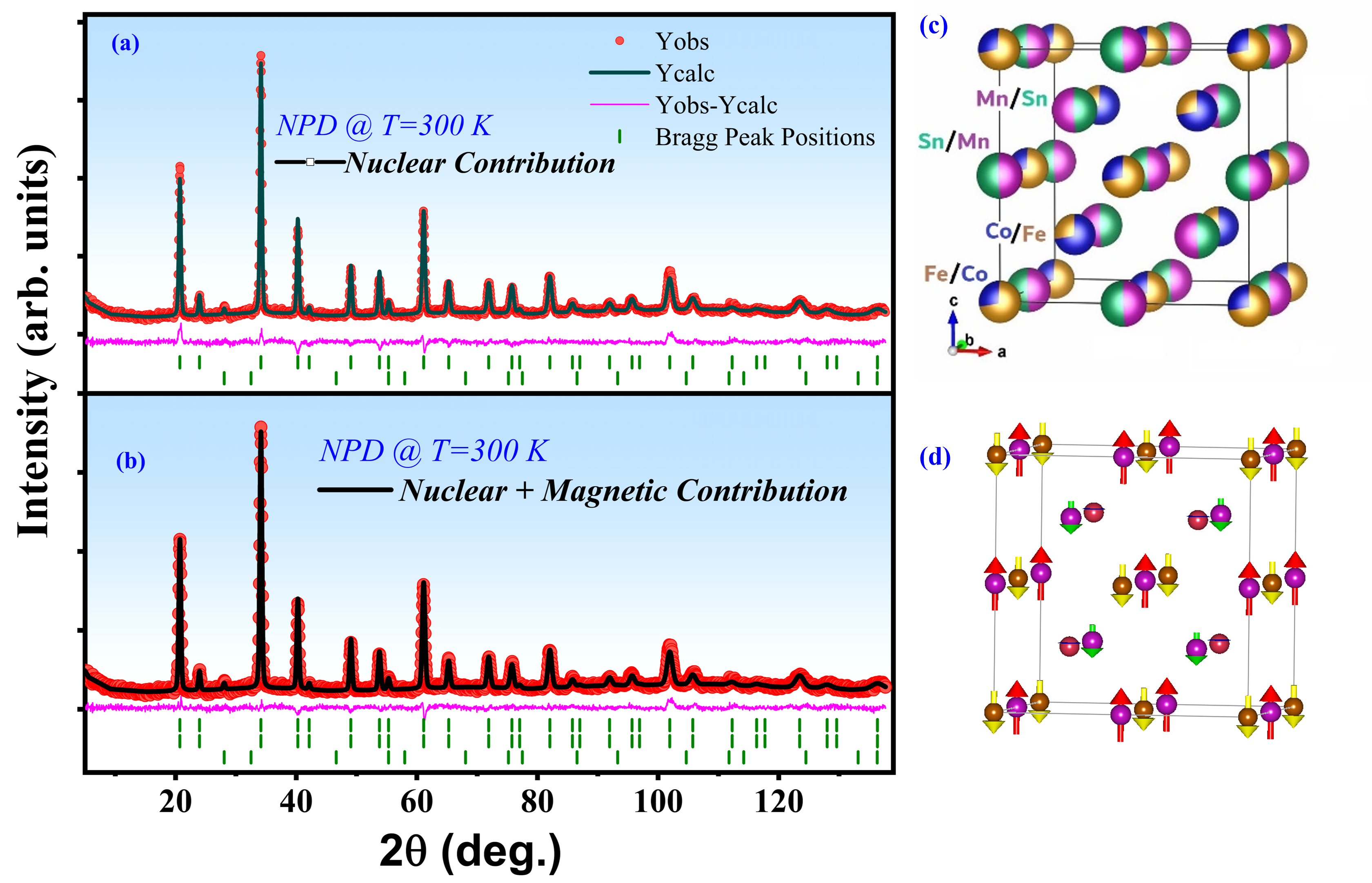}
\caption{(a) Experimentally observed (circles) measured on PD-2 diffractometer and calculated (solid lines through the data points) Neutron Powder diffraction (NPD) patterns for CoFeMnSn at 300 K. (b) NPD pattern at 300 K with magnetic phase contribution. (c) A schematic of the crystallographic unit cell with the relative distribution of the atoms over the four crystallographic sites (d) Schematic magnetic structure for CoFeMnSn. }
\label{fig:cd604}
\end{figure}

In CFMS, all its constituents (Co, Fe, and Mn) are the d-block elements of the periodic table. Due to their close proximity and similar X-ray scattering factors, XRD often does not provide the correct information regarding the atomic disorder within the crystal structure. Neutron diffraction, however, is more capable of distinguishing site disorder, as it can differentiate between atoms that have similar X-ray scattering factors yet exhibit significant differences in coherent scattering amplitudes. This makes neutron diffraction a more sensitive technique for identifying atomic site disorder in compounds such as CFMS compared to XRD. The room temperature neutron diffraction pattern is presented in figure \ref{fig:cd604} (a).
Usually, Heusler compounds possess disorder structure, and very common disorders are A2 type and B2 type. In an A2-type disorder, the atoms X, X', Y, and Z are randomly distributed throughout the crystal lattice. In contrast, a B2-type disorder involves selective random mixing, where Y and Z atoms, as well as X and X' atoms, mix within specific crystallographic positions, namely the 4a, 4b, 4c, and 4d sites. In an A2-type disorder, both superlattice reflections are missing, whereas only the (200) peak is present in a B2-type disorder. However, for CFMS, both the superlattice peaks are present in the neutron diffraction data recorded at 300 K. These peaks signify the formation of an ordered structure. We have performed the Rietveld refinement of the diffraction pattern measured at 300 K is shown in figure \ref{fig:cd604}. 2 with space group F$\overline{4}$3m. The large differences in the neutron scattering lengths of Co (2.490 fm), Fe (9.450 fm), Mn (-3.730 fm), and Sn (6.225 fm) allow us to determine relative site occupancies of the element over the four crystallographic sites 4a, 4b, 4c and 4d. Distribution of the Co/Fe atoms over the 4c/4a sites and Mn/Sn atoms over 4b/4d sites have been found shown in figure \ref{fig:cd604}. The refined parameters, along with the antisite disorder, are shown in Table II. 
\begin{table}[ht]
\caption{Site occupancies and site averaged magnetic moments of CoFeMnSn determined from the Rietveld analysis of the RT neutron diffraction pattern.}
\begin{ruledtabular}
\begin{tabular}{ccccc}
Site & Atoms &   Atoms Occupancies& $M_{av}(\mu_B) $\\ \hline
4c (1/4,1/4,1/4)& Co & 0.73(1)& 0.32(4) \\ 
& Fe & 0.27(1) \\ \hline
4a(0,0,0)& Fe & 0.73(1)& -1.23(3) \\ 
& Co & 0.27(1) \\ \hline
4b (1/2,1/2,1/2)& Mn & 0.48(7)& 2.87(2) \\ 
& Sn & 0.51(7) \\ \hline`
4d (3/4,3/4,3/4)& Sn & 0.48(7)& -0.95(2) \\ 
& Mn & 0.51(7) \\ 

\end{tabular}
\end{ruledtabular}
\end{table}

The Bragg peak at $\sim$ 28 deg. (denoted by the symbol $\#$ reveals the presence of a minor secondary phase of MnO ($\sim$ 0.6 wt$\%$).
The compound CFMS has been found to be in the magnetically ordered state at 300 K. This magnetic ordering is further validated through magnetization measurements, which established the presence of magnetic alignment even at room temperature. The additional intensities at the top of the nuclear Bragg peaks, especially at $\sim $ 20\textdegree  and 34\textdegree, reveal FM/FIM ordering in the compound.

\subsection{Isothermal Magnetization Study}
\begin{figure}[ht]
\centering
\includegraphics[width=1.1\linewidth]{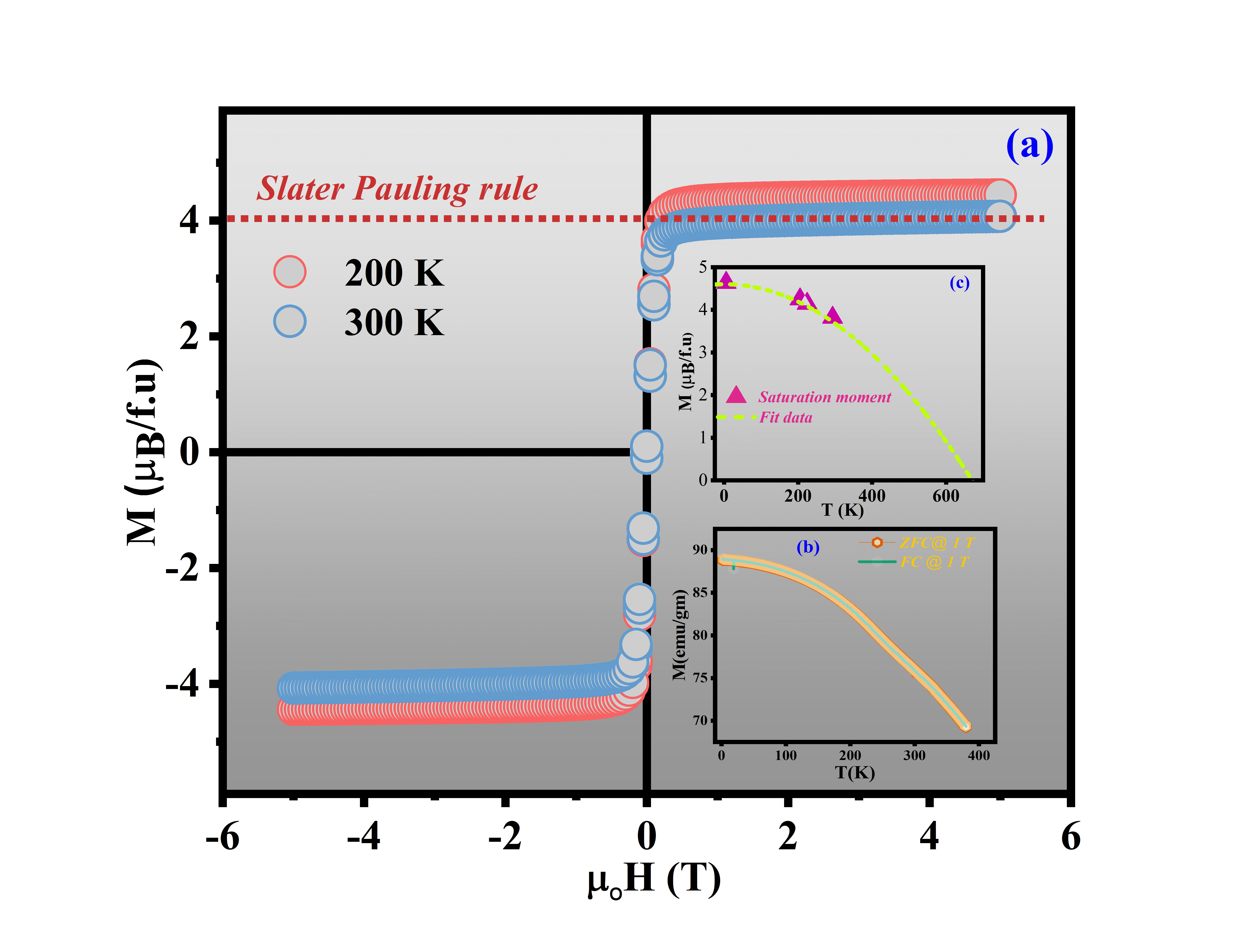}
\caption{Isothermal magnetization (M(H)) curve recorded at 200K and 300 K, dotted line shows the predicted value of $m_s$ from Slater Pauling rule. (b) Thermomagnetic data at 1 T in ZFC and FC condition. (c) This shows the fitting for estimated Curie temperature between $M_s$ and T.}
\label{fig:cd605}
\end{figure}

To understand the magnetic properties of the system, both isothermal magnetization \textit{(M(H))} (as shown in the figure \ref{fig:cd605} (a)) and temperature-dependent magnetization \textit{(M(t))} (as shown in the inset of the figure \ref{fig:cd605} (b) were measured. The \textit{M(H)} curve showed the saturation behavior, indicating that the system is ferromagnetic and magnetic moment decreased as the temperature increased from 2 K to 300 K. The value of saturation magnetization \textit{$(M_s)$} was observed  4.04 $\mu_B/f.u$.  for CFMS. All Heusler compounds adhere to a governing Slater Pauling (S-P) rule. According to this rule, the total magnetic moment for a Heusler alloy can be calculated using the formula m = ($N_V$ - 24) $\mu_B/f.u$. In this equation, m represents the total magnetic moment, and $N_V$ signifies the total valence electron count (VEC) of the compound \cite{graf2010heusler}. For the present system, the total VEC is 28. Therefore, the calculated value of  \textit{$(M_s)$} would be 4 $\mu_B/f.u.$, Which is in good agreement with the experimentally observed value. A noninteger value of m in the Heusler alloys suggested that the systems are half-metallic or spin-gapless in nature. 
Furthermore, the temperature-dependent magnetization data was recorded in both conditions zero field cooled (ZFC) and field cooled (FC) from 2K to 300 K after applying an external magnetic field of 1 T. To determine the Curie temperature $(T_c)$, this equation is used, $M_s= M_0[1-(T/T_c)^2]^{1/2}$. The estimated $(T_c)$ is 660 K, as shown in the inset of figure\ref{fig:cd605} (c); it is comparable with the experimental value \cite{gupta2023spin} 

\subsection{Transport properties} 
\begin{figure}[ht]
\centering
\includegraphics[width=1.0\linewidth]{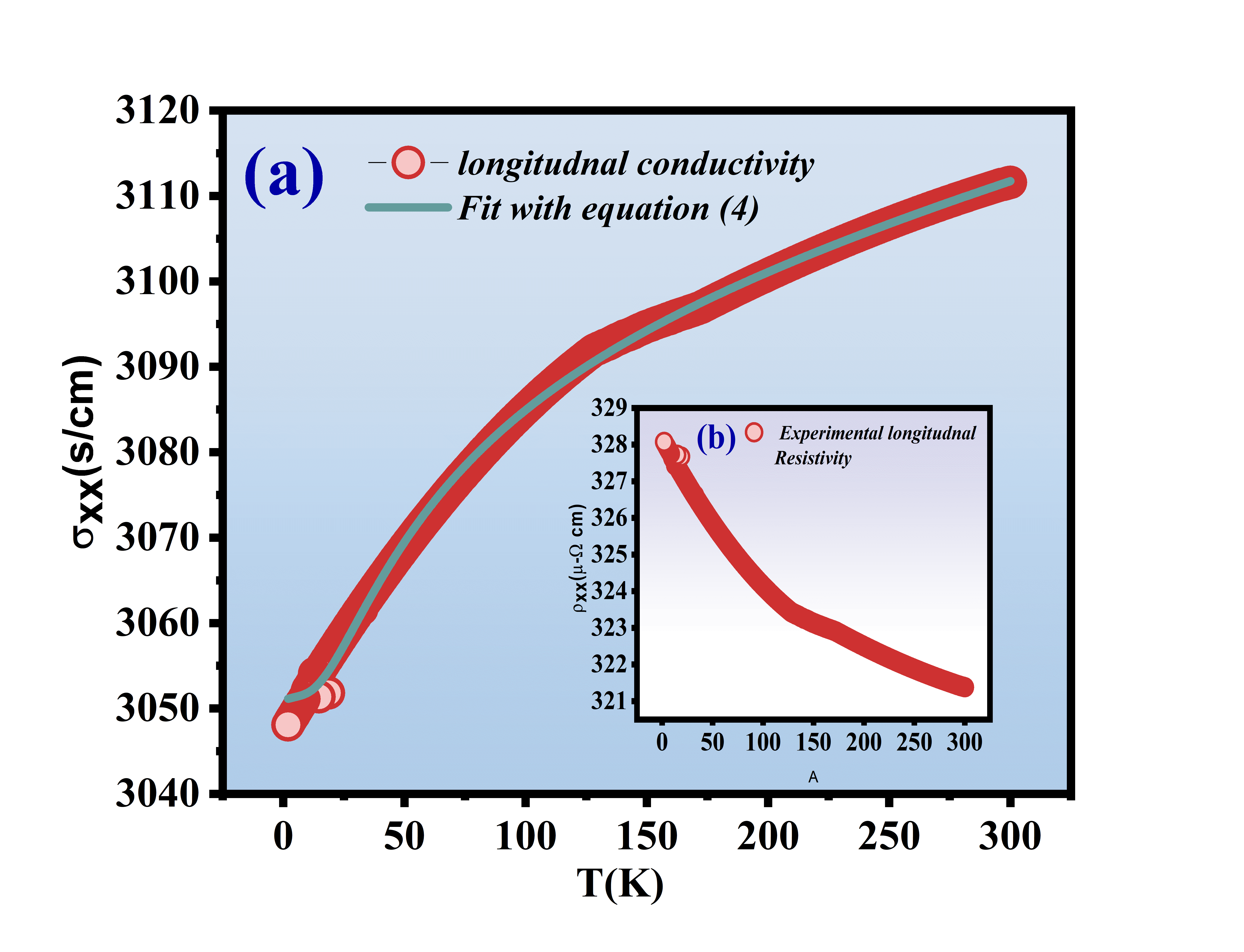}
\caption {Temperature dependent longitudinal electrical conductivity of the system fitted with the two carrier model using equation \ref{606} (in the main graph) (b) Inset showing the temperature variation of electrical resistivity.}
\label{fig:cd606}
\end{figure}

\begin{figure}[ht]
\centering
\includegraphics[width=1.0\linewidth]{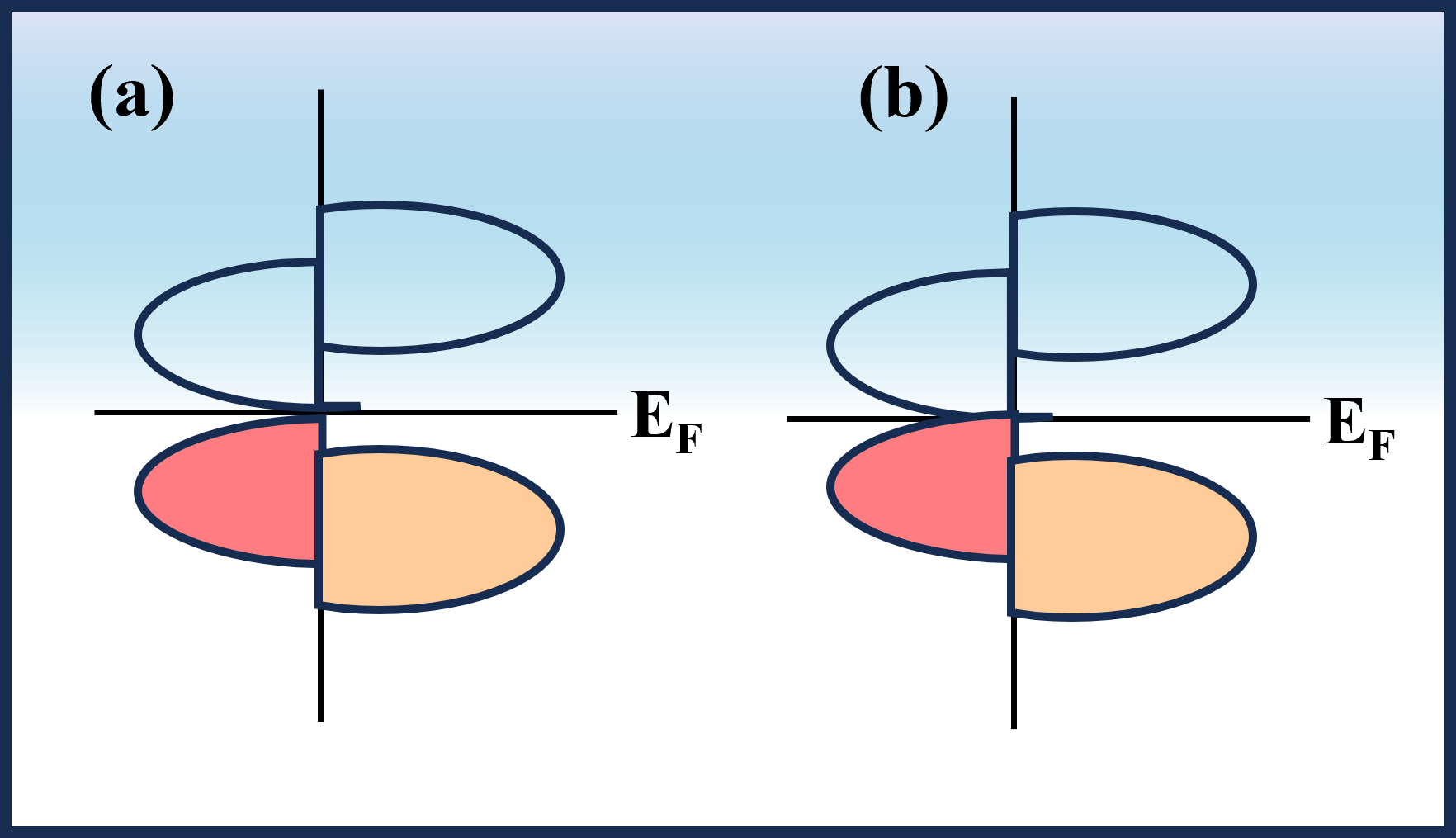}
\caption {This represents the density of a (a) Spin gapless semiconductor (g=0) (b) disordered Spin gapless semiconductor (g>0). }
\label{fig:cd606b}
\end{figure}

  Figure \ref{fig:cd606} shows the temperature-dependent longitudinal conductivity  ($\sigma_{xx}$)(in the main graph) and longitudinal resistivity ($\rho_{xx}$ (in the inset) from 2 K to 300 K of the system. It shows that resistivity decreases as the temperature increases, indicating the negative temperature coefficient of the resistivity. It shows a non-metallic behavior, and the minor dependency on temperature eliminates the traditional semiconducting behavior. A similar kind of behavior is also observed for other Heusler alloys, which exhibit spin-gapless semiconductors (SGSs) behavior. In Spin gapless semiconductors, one spin channel is gapless while the other spin channel behaves like a semiconductor, as shown in Figure \ref{fig:cd606b} (a). The resistivity of SGSs decreases linearly with temperature, unlike a typical semiconductor in which resistivity has exponential temperature dependence. Additionally, the range of $\sigma_{xx}$ usually observed for ordered SGSs falls between 3800 – 4500 S/cm. For instance, CoFeTiSn shows $\sigma_{xx}$  values ranging from approximately 2400 to 2650 S/cm \cite{xia2022magnetic}, for  CoFeCrGa from about 3130 to 3240 S/cm) \cite{bainsla2015origin}, and for CoFeMnSi from around 2940 to 2980 S/cm \cite{bainsla2015spin}. The conductivity of all these systems exhibits only a weak dependence on temperature. 
Few SGSs Heusler alloys have disordered structures,  which affects their magnetic and transport properties as observed for Mn$_2$CoAl \cite{shahi2022antisite}. In the case of disordered SGSs, instead of gapless behavior, the overlapping in density of states is observed at the fermi level, as shown in figure\ref{fig:cd606b} (b). In order to explain the temperature dependence of resistivity of the CFMS system, we have adopted the two-carrier model given by Kharel et al \cite{kharel2015magnetism}. This model accounts for the conductivity behavior of various spin-gapless and spin-narrow-gap semiconductors and disordered spin-gapless semiconductors. For the present study, the resistivity data is fitted with a given equation that incorporates contributions from both the gapless channel $(\sigma_{SGS})$ and the semiconducting channel $(\sigma_{SC})$. 

\begin{equation}
\sigma_{xx}(T)= \sigma_{SGS}+\sigma_{SC}
\end{equation}

\begin{equation}
\sigma_{xx}(T) =\sigma_0\left[1+2ln(2)\frac{K_BT}{g}\right]+\sigma_{SC0}exp\left(\frac{-\Delta E}{K_BT}\right)
   \label{606}
\end{equation}

The model's parameters include the zero-temperature conductivity of the gapless component $\sigma_0$, the band overlap parameter (g), Boltzmann’s constant ($k_B$), the semiconducting contribution at zero temperature ($\sigma_{SC0}$), and the carrier activation energy ($\Delta E$). The fitting yielded the value of $g \approx 0.5$meV, indicating a very small overlap of states instead of truly gapless condition (g=0).

\subsection{Hall Resistivity}
\begin{figure*}
\includegraphics[width=16cm,height=9.5cm]{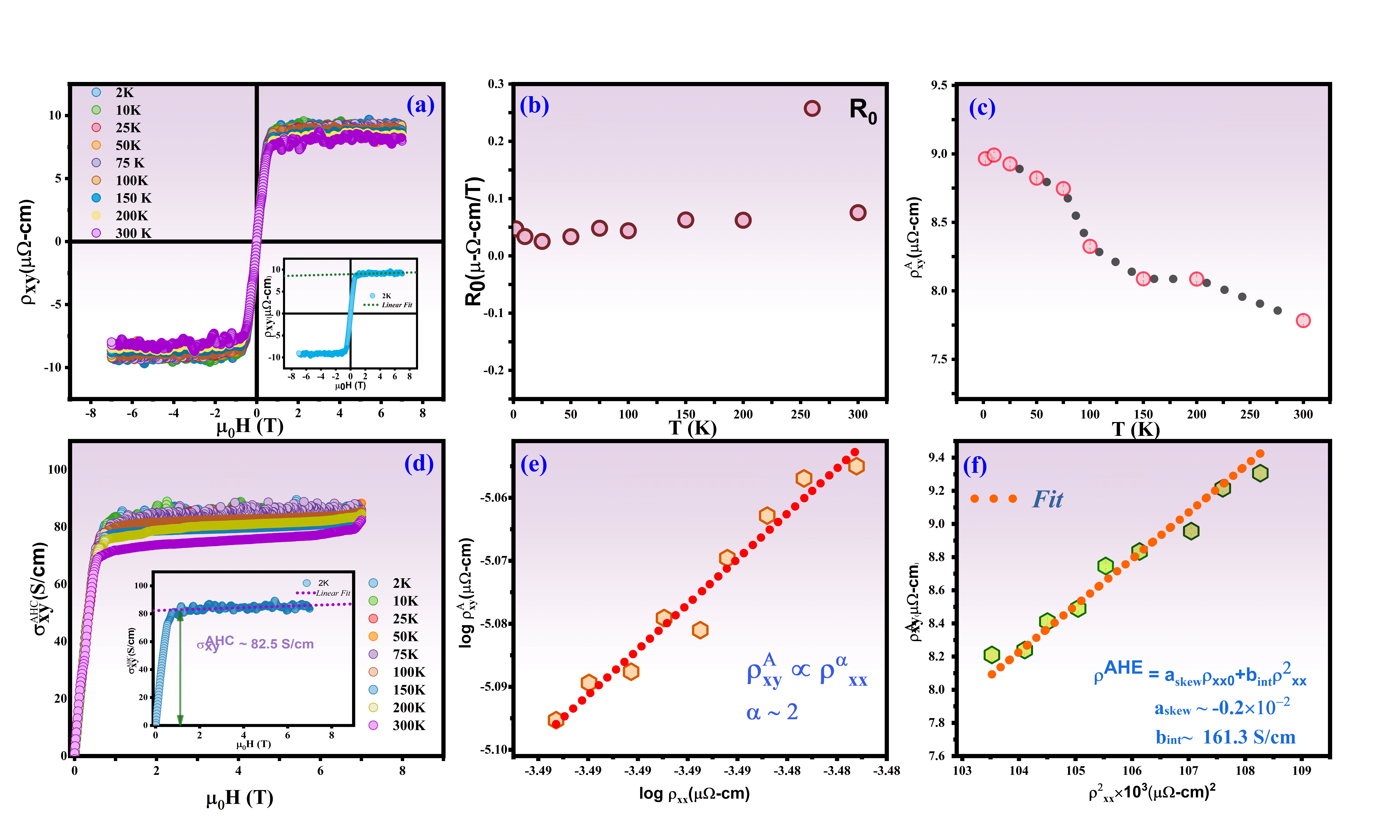}
\caption{(a) Field dependent Hall resistivity ($\rho_{xy}$) at various temperatures (Inset showing the linear fit of $\rho_{xy}$ at the high field region) (b) Ordinary Hall coefficient vs. temperature (c) Temperature-dependent value of Anomalous Hall resistivity extracted using equation \ref{608}. (d) Field-dependent Anomalous Hall conductivity (AHC) at different temperatures (Inset showing the zero-field extrapolation of high-field Hall conductivity data at 2K). (e) Scaling relation used in logarithmic scale plot between $\rho^A_{xy}$  and  $\rho_{xx}$ (f) $\rho^A_{xy}$  vs.  $\rho_{xx}$ curve  Fitted using equation \ref{610}}
\label{fig:cd607}
\end{figure*}

To understand the effect of antisite disorder on the transport properties of the CFMS system, the Hall resistivity ($\rho_{xy}$) of the system was measured for different temperature ranges between 2 K to 300 K as shown in figure \ref{fig:cd607} (a). The behavior of $\rho_{xy}$ is analogous to the M (H) behavior, suggesting the presence of AHE in the system. Now, the total $\rho_{xy}$ is explained by the following equation:
\begin{equation}
\rho_{xy}= R_o H + R_s M_s
\label{607}
\end{equation}

Here, $R_o$, $R_s$  is the ordinary, anomalous Hall coefficient respectively and $M_s$ is the magnetization of the system. The additional term $R_s M_s$, the contribution from the AHE, is due to the magnetization of the system. In order to separate the contribution from the anomalous part, the contribution coming from the ordinary part is subtracted from the total $\rho_{xy}$ as described in equation (\ref{607}). To do so, we began by performing the linear fitting in a higher field region (greater than 1T) as shown in the inset of figure \ref{fig:cd607} (a). After linear fit the data, the slope and the intercepts provide the values of $R_o$ and $R_sM_s$  respectively. The magnitude of AHE was 8.4 $\mu\Omega-cm$ at 2K, shown in the inset of figure \ref{fig:cd607} (a). The anomalous Hall conductivity $(\sigma_{xy}^{AHC})$ of the system is calculated by using the following equation:

\begin{equation}
    \sigma_{xy}^{AHC}=  \frac{\rho_{xy}}{ (\rho_{xy}^2+\rho_{xx}^2 )}
    \label{608}
\end{equation}

Figure \ref{fig:cd607} (d) shows the $\sigma_{xy}^{AHC}$  vs.H from 2 K to 300 K. The value of AHC at 2 K is 83.5 S/cm is observed. A similar value of AHC is found for other systems also \cite{mishra2022investigation,nag2022cofevsb}. Particularly here the experimentally observed value of AHC is comparable with the reported by Xia et.al.\cite{xia2023structural} viz; 87 S/cm. However, Gupta et al.\cite{gupta2023spin} reported a considerably smaller value, i.e., 53 S/cm at 5 K. The lower value observed by Gupta et. al. was due to the fact that compared to half-metallic ferromagnets (HMFs) and ferromagnetic metals, SGSs magnetic Heusler compounds exhibit a lower AHC. 
Understanding of the mechanism involved in AHE in the system is crucial for applying them in spintronics. Hence, to gain insight into the AHE and its underlying mechanism, both intrinsic and extrinsic, we have used a scaling model. 
\begin{equation}
    \rho_{xy} \propto \rho_{xx}^\alpha
    \label{609}
\end{equation}

These scaling models help in determining the separate contributions coming from extrinsic and intrinsic mechanisms as well as determining the dominating mechanism in AHE. In equation (\ref{609}) if $\alpha$ = 1 then the extrinsic mechanism is solely responsible for the AHE in the systems. If $\alpha$ =2, then it suggested that both intrinsic as well as side jump are responsible for the presence of AHE in the systems. The combined contribution of intrinsic mechanism and side jump cannot be further extinguished. For the present system, the value of $\alpha$ = 2 after plotting the equation (\ref{609}) on a logarithmic scale. This suggested that the intrinsic mechanism is dominating in the present system. Furthermore, the following scaling relation is used to determine the separate contributions from extrinsic and intrinsic mechanisms. 
\begin{equation}
  \rho_{xy}^{AHE}= a\rho_{xx0}+b\rho_{xx}^2  
  \label{610}
\end{equation}

Here, $a$ denotes the part that comes from extrinsic skew scattering, and $b$ shows the parts that come from intrinsic and side jump scattering. After fitting the relation given above (\ref{610}), the value of $a$ and $b$ is -0.002, 161.3 S/cm, respectively, as shown in figure \ref{fig:cd607} (f). The experimental value of observed AHC (viz 83.5 S/cm) is half of the theoretically obtained value after fitting (viz 161.5 S/cm). The disparity occurs because the intrinsic mechanism has an opposite nature to the extrinsic mechanism, as shown by the negative sign of a \cite{nag2022cofevsb}.

\subsection{Thermoelectric Power}

\begin{figure}[ht]
\centering
\includegraphics[width=1.0\linewidth]{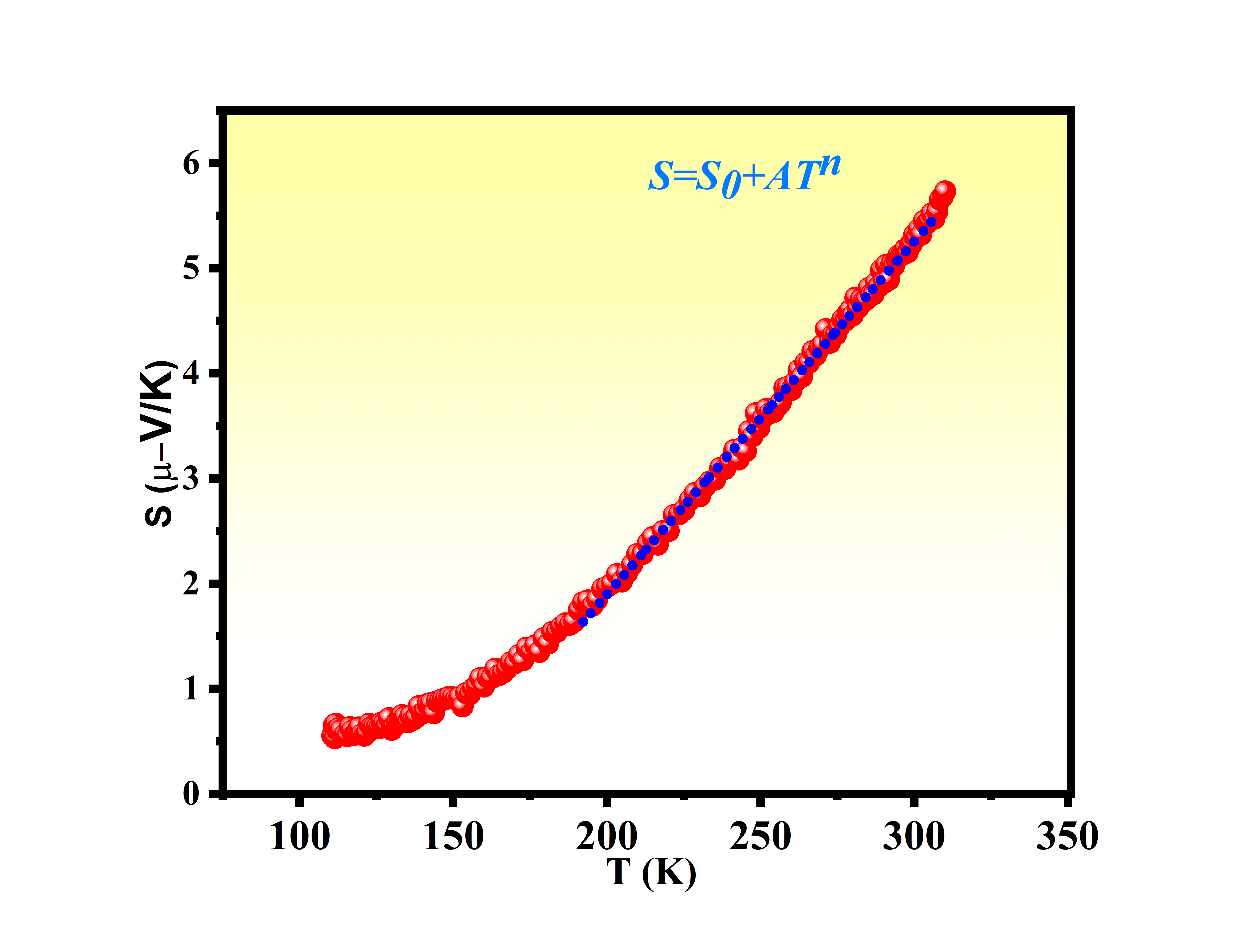}
\caption{Temperature dependent Seebeck coefficient (S) and the fitting to the S vs. T illustrates the linear behavior (blue solid line).}
\label{fig:cd608}
\end{figure}

Figure \ref{fig:cd608} shows the temperature variation of the Seebeck coefficient (S) from $100 K\leq T \leq 300 K$. The positive value of S indicates that the holes are the dominant carrier in these compounds. This is also in accordance with the Hall data also confirmed that the holes are the dominating charge carriers. Seebeck coefficient shows the linear variation with temperature. The value of S attains 6 $\mu-V/K$ at room temperature, comparable with other SGS Heusler compounds. Usually, a low value of S is observed for other SGSs as well \cite{shahi2022antisite, chanda2022emergence,rani2019spin,ouardi2013realization}.  The linear variation of S suggests a dominant contribution of diffusion thermopower as observed for other Heusler compounds\cite{nag2022cofevsb,nag2023nontrivial,nag2023topological,dixit2024raman}. However, in small or gapless semiconductors, $E_F$  can switch to the valence band (VB) or conduction band (CB) in one of the spin channels (with a very small gap) due to excitation or impurity states. This leads to a low S value and a slow linear variation.  To get an idea about the estimated  $E_F$ , S-data is fitted with the equation $S_d=S_0+aT$ in the high-T regime, where $S_d$ is the diffusion thermopower, $S_0$ is a constant given by,
\begin{equation}
 S_0=  (\pi^2 k_B^2)/(3eE_F )   
 \label{611}
\end{equation}

After fitting the above equation, the value of $E_F$ comes out to be 0.7 meV for the CFMS system, which is in accordance with the other SGSs Heulser alloys. 

\section{Conclusion} 
This study explores the impact of antisite disorder on various properties of the quaternary Heusler alloy CoFeMnSn. Although this alloy is classified as a spin gapless semiconductor (SGS), the presence of antisite disorder modifies its characteristics, leading to gapless semiconducting behavior. The investigation of structural, magnetic, and transport properties is conducted using several experimental techniques. Structural characterization is carried out using XRD and NPD analyses, both of which confirmed the existence of antisite disorder among the constituent elements of the system. The XMCD data revealed that the magnetic moments of the elements are aligned in parallel, indicating ferromagnetic ordering within the material. Magnetization studies further supported the ferromagnetic nature of the alloy, with a saturation magnetization of approximately 4 $\mu_B$/f.u. The Curie temperature for CoFeMnSn is estimated to be around 660 K. The carrier transport in the system is found to be governed by the two-carrier model, and the non-zero value of the parameter g indicates the overlapping states in one of the spin channels. The transverse resistivity shows the presence of AHE in the material. The AHC is calculated to be 83.5 S/cm at 2 K, and the antisite disorder was observed to enhance the AHC, signifying a significant impact on the electronic transport properties.
Additionally, the thermopower of the alloy is measured to be 5.7 $\mu$-V/K at room temperature. The combination of low thermopower, specific resistivity behavior, and the relatively low AHC value aligns with the typical features of a spin-gapless semiconductor, though slightly modified due to the presence of disorder. This study demonstrates that the controlled introduction of antisite disorder can effectively tune the electronic and magnetic properties of CoFeMnSn, positioning it as a promising material for advanced spintronic application.

\begin{acknowledgments}

The authors are grateful to the central instru-
ment facility of IIT(BHU) for providing the SEM, 
and the MPMS facility to execute magnetic measure-
ments.
\end{acknowledgments}

\bibliography{CFMS}

\end{document}